\documentclass[%
 reprint,
 amsmath,amssymb,
 aps,
prl]{revtex4-1}

\usepackage{subfigure}
\usepackage{color}
\usepackage{graphicx}
\usepackage{dcolumn}
\usepackage{bm}

\begin{document}

\preprint{APS/123-QED}

\title{Topological Phase Transition driven by Infinitesimal Instability:\\ Majorana Fermions in Non-Hermitian Spintronics}

\author{Nobuyuki Okuma}
\email{okuma@hosi.phys.s.u-tokyo.ac.jp}
\author{Masatoshi Sato}
\affiliation{%
 Yukawa Institute for Theoretical Physics, Kyoto University, Kyoto 606-8502, Japan
}%

%
\date{\today}
\begin{abstract}
Quantum phase transitions are intriguing and fundamental cooperative phenomena in physics. 
Analyzing a superconducting nanowire with spin-dependent non-Hermitian hopping, 
we discover a topological quantum phase transition driven by infinitesimal cascade instability.
The anomalous phase transition is complementary to the universal non-Bloch wave behavior of non-Hermitian systems.
We show that an infinite small magnetic field drastically suppresses the non-Hermitian skin effect, inducing a topological phase with Majorana boundary states. 
Furthermore, by identifying the bulk topological invariant, we establish the non-Hermitian bulk-boundary correspondence that does not have a Hermitian counterpart.
We also discuss an experimental realization of the system by using the spin-current injection to a quantum wire.

\end{abstract}

\pacs{}
\maketitle
Recently, non-Hermitian Hamiltonians \cite{Hatano-Nelson,Bender-98,Bender-02,Bender-review,Konotop-review,Christodoulides-review,Feng-review,Alu-review} have attracted much interest in various fields such as open systems \cite{Konotop-review,Christodoulides-review}, optics \cite{Malzard-15,Mochizuki-16, Xiao-17, St-Jean-17,Harari-18,Bandres-18,Feng-review,Alu-review,ozawa_topo_photo}, disordered (and correlated) systems \cite{Kozii-17,Zyuzin-18,Yoshida-prb18,Moors-19,Yoshida-19,Hamazaki-18}, and quantum critical phenomena \cite{KAU-17,Nakagawa-18,Xiao-18,Dora-18}. 
Among them, topological properties of such Hamiltonians have been extensively investigated both in gapped \cite{Rudner-09,Hu-11,Esaki-11,Sato-12,Poli-15,Lee-16,Weimann-17,Leykam-17,Obuse-17,Mochizuki-16, Xiao-17, St-Jean-17,ozawa_topo_photo,Harari-18,Bandres-18,Shen-18,KAKU-18,KSU-18,YW-18,YSW-18,Kunst-18,Borgnia-19,Yokomizo-19,Lieu-18,Liu-19,Gong-18,KHGAU-19,KSUS-18,CHLee-1809,CHLee-1810} and gapless phases \cite{Yoshida-19,Kozii-17, Zyuzin-18, Moors-19,Malzard-15,Xu-17,Cerjan-18,Cerjan-18-exp, Zhen-15,Gonzalez-16,Gonzalez-17,Molina-18,Chernodub-17,Chernodub-19,Zyuzin-19,Zhou-18-exp,CSBB-18,Okugawa-19,Budich-19,Yang-19,Zhou-19,Wang-19,Luo-18,Lee-18-tidal,KBS-19,Xiaosen-19}, and a lot of essential differences from the Hermitian cases have been pointed out. 
For instance owing to the complex nature of the energy spectrum, there are several distinct definitions of energy gaps, which amplify the possibility of topological phases \cite{Gong-18,KHGAU-19,KSUS-18}.

Although the non-Hermitian physics under the periodic boundary condition (PBC) can be investigated by using mathematical tools developed in the Hermitian physics \cite{Hasan-Kane,XLQ-SCZ,Schnyder-08, Kitaev-09,Ryu-10, Freed-Moore,Gomi-17,Morimoto-Furusaki,Shiozaki-14,Schnyder-Ryu-review},
it is not easy to treat them under the open boundary condition (OBC) because of the non-Bloch wave behavior \cite{Hatano-Nelson}.
For instance, the phase diagrams under the OBC are different from those under the PBC in several non-Hermitian models \cite{YW-18,YSW-18,KSU-18}, which obscures the conventional bulk-boundary correspondence.
Thus far, the non-Hermitian bulk-boundary correspondence has not been established except for several attempts \cite{YSW-18,YW-18,Kunst-18,Borgnia-19,Yokomizo-19}.

In this Letter, we construct and analyze a simple non-Hermitian lattice model that describes a one-dimensional $s$-wave superconductor with spin-dependent asymmetric hopping.
Although this model is topologically nontrivial under the PBC, the SU(2) imaginary gauge transformation reveals that 
the system under the OBC does not show any topological boundary modes.
Interestingly, however, we find that this mismatching is drastically remedied by an infinitesimal transverse magnetic field in the thermodynamic limit.
Performing the numerical diagonalization with a small magnetic field, we find the missing Majorana boundary modes, which are protected by the $\mathbb{Z}_2$ topological invariant. 
This finding establishes the presence of the non-Hermitian bulk-boundary correspondence that has no analog in the Hermitian physics. Finally we also discuss an experimental realization by using the spin current injection to a quantum wire.

\begin{figure}[]
\begin{center}
　　　\includegraphics[width=7cm,angle=0,clip]{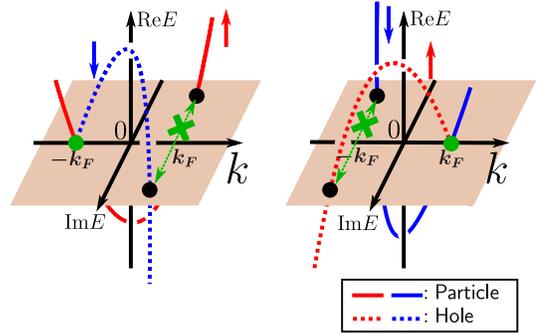}
　　　\caption{Schematic picture of the Bogoliubov band structure in the normal state ($\Delta=0$) with the non-Hermitian spin-orbit interaction. $k=\pm k_F$ are the Fermi momenta. When nonzero $\Delta$ is introduced, electrons with sufficiently large dissipation ($\Gamma>\Delta$ at black dots) cannot form the Cooper pair due to the distance in the complex-energy space (green dotted arrows), while those with no dissipation can (green dots). The spin- and momentum-dependent pairing is introduced as a source of the topological phase transition. In the real-space picture with open boundaries, however, there is a subtlety due to the non-Hermiticity (see main text for details).}
　　　\label{fig1}
\end{center}
\end{figure}

{\it Periodic boundary condition}.\textemdash
In order to grasp a rough idea,  
we first analyze the infinite lattice system under the PBC, where the momentum-space picture is useful.
For Hermitian systems, the Majorana fermions are known to appear on boundaries of a spinless $p$-wave superconductor \cite{Kitaev2001}, though it is not experimentally relevant thus far.
Instead of the direct realization, several schemes have been proposed to effectively create the spinless Cooper pairing \cite{Sato-Ando,Sato-Fujimoto,Lutchyn,O-R-vO}.
For example, in a quantum wire with the Rashba spin-orbit interaction and the Zeeman magnetic field, the spin degree of freedom is frozen due to the spin-momentum locking \cite{Lutchyn,O-R-vO}.
In this paper, we use the spin-momentum locked dissipation to realize a similar spinless situation.

\begin{figure*}[]
\begin{center}
　　　\includegraphics[width=16cm,angle=0,clip]{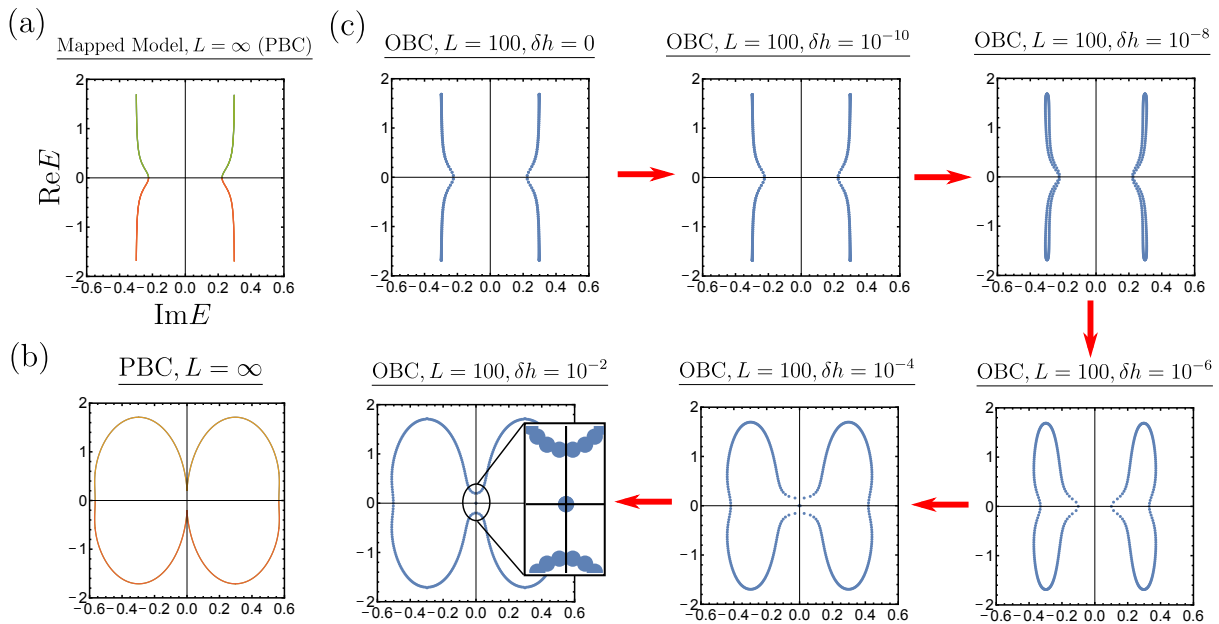}
　　　\caption{(a) Energy spectrum of the mapped model with the periodic boundary condition (PBC) obtained by using Eq. ($\ref{obcspectrum}$). It corresponds to the Hamiltonian ($\ref{latticemodel}$) with infinite sites under the open boundary condition (OBC). (b) Energy spectrum of the Hamiltonian ($\ref{latticemodel}$) with infinite sites under the PBC obtained by using Eq. ($\ref{pbcspectrum}$). (c) Energy spectrum of the Hamiltonian ($\ref{latticemodel}$) with 100 sites under the OBC for various small transverse magnetic fields. The result for $\delta h=0$ is well described by Eq. ($\ref{obcspectrum}$). For small but finite $\delta h$, on the other hand, the spectrum differs from Eq. ($\ref{obcspectrum}$), and two superposition states of Majorana fermions ($E=\pm4\times10^{-5}i$) appear in the cases with $\delta h=10^{-2}$ and $10^{-4}$. The model parameters are $t_+=1,t_-=0.7$ and $\Delta=0.2$. }
　　　\label{fig2}
\end{center}
\end{figure*}

Let us consider the following Hamiltonian:
\begin{align}
&H=H_N+H_{\Delta},\notag\\
&H_N=\sum_{k,\sigma_z=\pm}\left[-2t\cos k-\frac{i\Gamma}{2}(1+\sin k \sigma_z) \right]a^\dagger_{k,\sigma_z}a_{k,\sigma_z},\notag\\
&H_\Delta=\sum_{k}\left[\Delta a^\dagger_{k,\uparrow}a^\dagger_{-k,\downarrow}+\mathrm{H.c.}\right],\notag\\\label{momentumspace}
\end{align}
where $(a^{\dagger},a)$ are spin-$1/2$ fermionic (electron) creation and annihilation operators, and real parameters $t,\Gamma$ and $\Delta$ describe the kinetic energy, dissipation (loss for particles and gain for holes), and an $s$-wave gap function \footnote{Since the phase of gap function does not change the physics, the gap function is taken to be real.}, respectively.
The spin- and momentum-dependent dissipation is a non-Hermitian variant of spin-orbit interaction.
For $\Gamma>\Delta$, only the left-(right-)going electrons with up-(down-)spin can participate in the Cooper pairing (Fig. $\ref{fig1}$), and thus we obtain an effective spinless superconductor.
Diagonalizing the Hamiltonian, we have 
\begin{align}
H=\sum_{k,a=\pm}E_{k,a}\overline{\alpha}_{k,a}\alpha_{k,a},\label{pbcspectrum}
\end{align}
where $(\overline{\alpha},\alpha)$ are creation and annihilation operators of the Bogoliubov quasi-particles, and $E_{k,\pm}=\sqrt{\left[-2t\cos k-i\Gamma/2(1\pm\sin k) \right]^2+\Delta^2}$ are their energy dispersion [Fig.$\ref{fig2}$(b)]. Note that $\overline{\alpha}$ is no longer the Hermitian conjugate of $\alpha$ under the non-Hermiticity, while the conventional anti-commutation relation $\{\alpha_{k,a},\overline{\alpha}_{k',a'}\}=\delta_{kk'}\delta_{aa'}$ holds, and $\alpha$ annihilates the BCS vacuum $|0\rangle$ (see Supplemental Material (SM) or Ref. \cite{Yamamoto-19}).

In the real-space picture, Eq. ($\ref{momentumspace}$) can be written as a simple lattice model
\begin{widetext}
\begin{align}
H=\sum_{i,\sigma_z=\pm}\left[-t_{\sigma_z}a^\dagger_{i+1,\sigma_z}a_{i,\sigma_z}-t_{(-\sigma_z)}a^\dagger_{i,\sigma_z}a_{i+1,\sigma_z}-i(t_+-t_-)a^\dagger_{i,\sigma_z}a_{i,\sigma_z}\right]+\sum_{i}\left[\Delta a^\dagger_{i,\uparrow}a^\dagger_{i,\downarrow}+\mathrm{H.c.}\right],\label{latticemodel}
\end{align}
\end{widetext}
where $i$ is the site index, and $t_{\pm}=t\pm\Gamma/4$.
Note that the normal part of the Hamiltonian includes the non-Hermitian asymmetric hopping terms whose asymmetry depends on the $z$-component spin.
These hopping terms are regarded as those of a stacked Hatano-Nelson model \cite{Hatano-Nelson} with up and down spins.

{\it Open boundary condition}.\textemdash
Thus far, we have introduced the non-Hermitian spin-orbit interaction for the purpose of the realization of Majorana boundary states.
This proposal is based on the momentum-space picture, which corresponds to the PBC in real space.
In the presence of non-Hermiticity, however, extensive sensitivity of the energy spectrum to boundary conditions obscures the naive bulk-boundary correspondence.

Let us impose the OBC on the Hamiltonian ($\ref{latticemodel}$).
To consider the eigenvalues of Eq. ($\ref{latticemodel}$), we perform the SU(2) imaginary gauge transformation, which generalizes the imaginary gauge transformation \cite{Hatano-Nelson} used in the analysis of the Hatano-Nelson model:
\begin{align}
&a_{i,\uparrow}=\left(\sqrt{\frac{t_+}{t_-}}\right)^ib_{i,\uparrow}, a^\dagger_{i,\uparrow}=\left(\sqrt{\frac{t_+}{t_-}}\right)^{-i}b^\dagger_{i,\uparrow},\notag\\
&a_{i,\downarrow}=\left(\sqrt{\frac{t_+}{t_-}}\right)^{-i}b_{i,\downarrow}, a^\dagger_{i,\downarrow}=\left(\sqrt{\frac{t_+}{t_-}}\right)^{i}b^\dagger_{i,\downarrow},\label{gaugetransf}
\end{align}
where $(b^\dagger,b)$ are creation and annihilation operators in the new basis.
Under this transformation, the spin-dependent asymmetric hopping terms are mapped to the spin-independent symmetric ones:
\begin{align}
-\sqrt{t_+t_-}b^\dagger_{i+1,\sigma_z}b_{i,\sigma_z}+\mathrm{H.c.},
\end{align}
while the other terms are invariant.
Thus, the Hamiltonian ($\ref{latticemodel}$) is mapped to a conventional $s$-wave superconductor, apart from a constant dissipation term.
Although this transformation changes the eigenfunctions drastically, it does not change the eigenvalues because it is a similarity transformation.
Thus, the Hamiltonian ($\ref{latticemodel}$) has the same energy spectrum as that of the mapped $s$-wave superconductor.
This implies that the lattice model ($\ref{latticemodel}$) is topologically trivial under the OBC and has no Majorana boundary modes. 
The energy spectrum with a constant dissipation does not depend on the boundary condition in the thermodynamic limit as in the case of the Hermitian physics, and it is calculated as
\begin{align}
E_{\bm{k},a}=\sqrt{\left\{-2\sqrt{t_+t_-}\cos k-i(t_+-t_-)\right\}^2+\Delta^2},\label{obcspectrum}
\end{align}
which does not depend on the band index $a=\pm$ \footnote{Note that the imaginary gauge transformation and the above mapping is not consistent with the PBC.}.
The obtained spectrum is drastically different from Eq. ($\ref{pbcspectrum}$) [Fig.$\ref{fig2}$(a)].

At first sight, the above consideration seems to ruin the scenario of Majorana modes by the non-Hermitian spin-orbit interaction.
Interestingly, however, an infinitesimal perturbation resurrects this scenario as shown below.

{\it Phase transition driven by infinitesimal instability}.\textemdash
In the case of the OBC, asymmetry of the hopping induces the accumulation of eigenstates near boundaries. 
This non-Bloch wave behavior (non-Hermitian skin effect) stems from the exponentially growing form factor in Eq. (\ref{gaugetransf}), which breaks the conventional bulk-boundary correspondence.   
In the case of our model, however, this effect is drastically suppressed by adding the transverse Zeeman term
\begin{align}
H_{\mathrm{ex}}=\delta h\sum_{i}\left[a^\dagger_{i,\uparrow}a_{i,\downarrow}+\mathrm{H.c.}\right].\label{transversemagfield}
\end{align}
This term is not invariant under the SU(2) imaginary gauge transformation,
\begin{align}
H_{\mathrm{ex}}=\delta h\sum_{i}\left[\left(\frac{t_-}{t_+}\right)^i b^\dagger_{i,\uparrow}b_{i,\downarrow} +\left(\frac{t_+}{t_-}\right)^ib^\dagger_{i,\downarrow}b_{i,\uparrow}      \right],\label{infinitesimal}
\end{align}
and thus our model is no longer equivalent to the trivial $s$-wave superconductor even when $\delta h$ is very small.
Roughly speaking, this perturbation cannot be ignored if
\begin{align}
|\delta h|\left(\frac{t_+}{t_-}\right)^{L/2} &\gtrsim \mathcal{O}(t,\Gamma,\Delta)\notag\\
\Leftrightarrow |\delta h| &\gtrsim \alpha_1e^{-\alpha_2 L},\label{sizeeffect}
\end{align}
where $L$ is the system size, and $\alpha_1,\alpha_2>0$ are constants.
After taking the thermodynamic limit, the infinitesimally small perturbation drastically changes the energy spectrum from the unperturbed one. In other words, the order-of-limits changes the physics:
\begin{align}
\lim_{\delta h\rightarrow0}\lim_{L\rightarrow\infty}\neq\lim_{L\rightarrow\infty}\lim_{\delta h\rightarrow0}.
\end{align}
Similar high sensitivity of eigenvalues to the pertubation is also discussed in 
mathematics \cite{pseudoeigen}.
It is important to note that nonlocal perturbations such as the coupling between two ends destroy the OBC nature, while local perturbations such as the transverse magnetic field in our model can preserve it. This difference leads to the emergence of the topological boundary modes. In general, such an infinitesimal instability against a local perturbation occurs when (i) the generalized Bloch Hamiltonian \cite{Yokomizo-19} is divided into two different eigen sectors of a unitary symmetry operator, (ii) the two sectors are energetically degenerate, but have different values that characterize the non-Hermitian skin effects,  and (iii) those two sectors are coupled via the perturbation (see details for SM) \footnote{In our model, the unperturbed Hamiltonian can be block-diagonalized with respect to the $z$-component spin.  }

The terms in Eq. (\ref{infinitesimal}) grow exponentially near boundaries, getting rid of the accumulated states of the skin effect. 
This implies that the OBC bulk spectrum 
would be close to the PBC one [Eq.($\ref{pbcspectrum}$)] in the presence of the perturbation.
In the following, we perform the numerical diagonalization to confirm this expectation.

{\it Numerical diagonalization}.\textemdash
We rewrite the Hamiltonian ($\ref{latticemodel}$) with the small perturbation ($\ref{transversemagfield}$) in the Nambu representation:
\begin{align}
H+H_{\mathrm{ex}}=\frac{1}{2}\sum_{i,j} \Psi^\dagger_{i}\mathcal{H}^{\mathrm{BdG}}_{i,j}\Psi_{j},
\end{align}
where $\Psi^\dagger_i=(a^\dagger_{i,\uparrow},a^\dagger_{i,\downarrow},a_{i,\uparrow},a_{i,\downarrow})$ is the Nambu spinor, and $\mathcal{H}^{\mathrm{BdG}}$ is the Bogoliubov-de Gennes (BdG) Hamiltonian matrix (explicit form in SM). Using the Nambu representation, we numerically calculate the energy spectra of the finite lattice system ($L=100$) for various transverse magnetic fields and plot them in Fig. $\ref{fig2}$ (c).
The model parameters are $t_+=1,t_-=0.7$ and $\Delta=0.2$.
$\Gamma$ is set to be larger than $\Delta$ in order to realize the spin-momentum locked Cooper pairing (Fig. $\ref{fig1}$). 

\begin{figure}[]
\begin{center}
　　　\includegraphics[width=6cm,angle=0,clip]{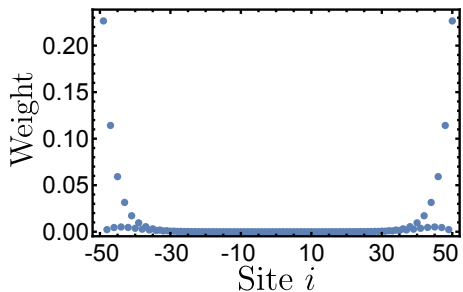}
　　　\caption{Weight function of a superposition state of the Majorana fermions ($E=-4\times10^{-5}i$). The model parameters are $t_+=1,t_-=0.7$, $\Delta=0.2$, $L=100$, and $\delta h=10^{-2}$.  }
　　　\label{fig3}
\end{center}
\end{figure}

In the absence of the magnetic field, the result is well approximated by Eq. ($\ref{obcspectrum}$). 
For $\delta h\gtrsim 10^{-8}$, the energy spectrum differs from Eq. ($\ref{obcspectrum}$), which is consistent with the value $(t_-/t_+)^{L/2}\simeq 2\times10^{-8}$ in Eq. ($\ref{sizeeffect}$). 
For $\delta h\gtrsim 10^{-4}$, we find two superposition states of Majorana fermions localized on two boundaries of the lattice system (Fig. $\ref{fig3}$), while the bulk spectrum surrounds the origin of the complex plane and has the similar shape as the spectrum under the PBC. As we expected, the small perturbation changes the spectrum into the topological one with Majorana fermions.

The Majorana fermions satisfy the non-Hermitian Majorana condition (see SM) \footnote{Since the non-Hermitian Majorana fermions also satisfy the important property that the creation and annihilation of particles are equivalent, they have the non-Abelian statistics as in the case of Hermitian one. All known exotic phenomena such as the fractional Josephson effect are expected to be observed.}:
\begin{align}
\overline{\gamma}_i=\gamma_i,
\end{align}
where $i=1,2$ denote the boundaries on which the Majorana fermions localize.
The effective theory of the two edges (1 and 2) are given by
\begin{align}
H_{\mathrm{Boundary}}&=\frac{i \epsilon}{2}\gamma_1\gamma_2=\epsilon (\overline{\alpha}\alpha-\frac{1}{2}),
\end{align}
where $\epsilon$ is the complex finite-size coupling, and $(\alpha,\overline{\alpha})$ are the fermion operators constructed from the Majorana fermions:
\begin{align}
\gamma_1=\alpha+\overline{\alpha},\quad\gamma_2=\frac{\alpha-\overline{\alpha}}{i}.
\end{align}
 In the present numerical calculation, the fermion energy $\epsilon$ takes the imaginary number ($\epsilon=-4\times10^{-5}i$).

{\it Non-Hermitian topological phase}.\textemdash
In the presence of the non-Hermitian skin effect, it is necessary to define the topological number by using the non-Bloch wave functions \cite{YSW-18,YW-18,Kunst-18,Yokomizo-19}.
In our case with small magnetic field, however, the numerical calculation indicates that the non-Bloch wave behavior reduces to the conventional Bloch one in the thermodynamic limit. 
%
In fact, it can be proven by combing the method in Ref. \cite{Yokomizo-19} and a symmetry consideration (see SM).
In the following, we identify the topological invariant that protects the Majorana zero mode in terms of the BdG Bloch Hamiltonian in PBC.

Let us consider the BdG Bloch Hamiltonian constructed from Eq. ($\ref{momentumspace}$)
\begin{align}
\mathcal{H}^{\mathrm{BdG}}_{k}=\left[-2t\cos k-\frac{i\Gamma}{2}\right]\hat{\tau}_z-\frac{i\Gamma}{2}\sin k \hat{\sigma}_z-\Delta\hat{\sigma}_y\hat{\tau}_y,\label{bdgrep}
\end{align}
where $\hat{\sigma}$s and $\hat{\tau}$s are the Pauli matrices in the spin and particle-hole space, respectively.
This Hamiltonian belongs to class D in the the Altland-Zirnbauer \cite{Altland-Zirnbauer} classification, and supports the particle-hole symmetry
\begin{align}
\hat{\tau}_x(\mathcal{H}^{\mathrm{BdG}}_{k})^T\hat{\tau}_x=-\mathcal{H}^{\mathrm{BdG}}_{-k}.\label{bdgbloch}
\end{align}
Note that the transpose in the charge conjugation is not equivalent to the complex conjugation for the non-Hermitian case \cite{chargeconjugation}.

The bulk band is not gapped in a usual sense since the bulk spectrum in Fig.\ref{fig1}(b)
is totally connected in the complex energy spectrum.
Therefore, the conventional class D topological invariant or its non-Hermitian variant is no longer well-defined; our obtained topological phase originates essentially from non-Hermiticity \cite{Gong-18}.
Hence we use another topological invariant intrinsic to non-Hermitian systems.
We propose the following $\mathbb{Z}_2$ invariant to characterize the present topological phase:
\begin{align}
(-1)^\nu=\frac{\mathrm{Pf}(\tau_x\mathcal{H}_{k=\pi})}{\mathrm{Pf}(\tau_x\mathcal{H}_{k=0})}\exp\left(-\frac{1}{2}\int_0^{\pi}dk\mathrm{Tr}[\mathcal{H}_k^{-1}\partial_k\mathcal{H}_k ]\right),
\end{align}
where $k=0,\pi$ are the time-reversal-invariant points, and the superscript ``BdG" is omitted.
The topological invariant is well-defined unless ${\rm det}{\cal H}_k= 0$ (i.e. $|\Gamma|=|\Delta|$).
The competition between $\Delta$ and $\Gamma$ determines the topological phase; the $\mathbb{Z}_2$ index is trivial for $|\Gamma|<|\Delta|$ and nontrivial for $|\Gamma|>|\Delta|$ (see SM). Thus, the present strong $\Gamma$ case, where the boundary modes exist, is topologically non-trivial, while the weak $\Gamma$ case, where the boundary modes are absent (see SM), is topologically trivial.

Although non-Hermitian topological phases whose gap-closing point is defined by $\det \mathcal{H}_k=0$ had been understood in the absence of boundaries, their topological nature under the OBC was unclear so far\cite{Gong-18}. Our results firstly establish the bulk-edge correspondence in this context, while whether it can be extended to the other Altland-Zirnbauer classes is a nontrivial open question.
The removability of the non-Hermitian skin effect under given symmetry would play an essential role to realize the bulk-edge correspondence.
 

{\it Spintronic application}.\textemdash
We finally discuss an experimental realization of the Hamiltonian ($\ref{latticemodel}$).
The nontrivial task is to implement the spin-dependent asymmetric hopping terms, or equivalently, the non-Hermitian spin-orbit interaction.
Although the full implementation of the $\sin k\ \sigma_z$ term seems to be difficult, 
we may introduce the essentially the same effect near the Fermi level, where the superconducting pairing occurs.
%
In order to introduce the spin-momentum locked effect near the Fermi level, we propose to a pure spin current injection to a quantum wire.
Under this non-equilibrium circumstance, modes with spin current $\sigma_z \partial E/\partial k$ opposed to the injected spin current would have a shorter life time by scatterings.
For sufficiently large imbalance of dissipation $\Gamma$, we obtain the situation in Fig. $\ref{fig1}$.
Another promising platform is ultra-cold atom systems.
The possible realization of the asymmetric hopping term has been theoretically proposed in Ref. \cite{Gong-18}, which would be generalized to our model with the spin degrees of freedom.

In summary, we have constructed and analyzed a simple non-Hermitian lattice model of an $s$-wave superconductor that realizes a novel topological phase.
The topological phase transition is driven by an infinitesimal external magnetic field.
We have also discussed an experimental realization in spintronics.
The present model provides the first concrete example of the non-Hermitian topological phase that does not have Hermitian counterparts.

\section*{Acknowledgements} We acknowledge many fruitful discussions with Ken Shiozaki, Takumi Bessho, and Ryusuke Hamazaki.
N.O. is grateful to Kohei Kawabata and Ken Mochizuki for introducing various topics about the non-Hermitian physics.
This work was supported by a Grant-in-Aid for Scientific Research on Innovative Areas ``Topological Materials Science" (KAKENHI Grant No. JP15H05855) from the Japan Society for the Promotion of Science (JSPS). N.O. was supported by KAKENHI Grant No. 18J01610 from the JSPS. M.S. was supported
by KAKENHI Grant No. JP17H02922 from the JSPS.


\begin{thebibliography}{93}
\bibitem [{cha()}]{chargeconjugation}%
  \BibitemOpen
  \href@noop {} {}\bibinfo {note} {The transpose in the particle-hole symmetry
  naturally arises when we anticommute the fermionic operators: $\sum_{i,j}
  h_{i,j}f_if_j$$\rightarrow \sum_{i,j} (-h_{i,j})f_jf_i=\sum_{i,j}
  (-h^T_{i,j})f_if_j$. In Hermitian physics, the transpose is equivalent to the
  complex conjugation, and the charge conjugation is antiunitary. In
  non-Hermitian physics, however, it does not hold. See Ref.~\cite{KSUS-18} for
  details.}\BibitemShut {Stop}%
\end{thebibliography}
%

\end{document}


\title{Supplemental Material for\\ ``Topological Phase Transition driven by Infinitesimal Instability: Majorana Fermions in Non-Hermitian Spintronics''}

\author{Nobuyuki Okuma}
\email{okuma@hosi.phys.s.u-tokyo.ac.jp}
\author{Masatoshi Sato}
\affiliation{%
 Yukawa Institute for Theoretical Physics, Kyoto University, Kyoto 606-8502, Japan
}%

\date{\today}
\maketitle

\section{Explicit form of the real-space BdG Hamiltonian}
For convenience, we here write down the explicit form of the Bogoliubov-de Gennes Hamiltonian in real space. The matrix elements are given by
\begin{align}
\mathcal{H}'^{\mathrm{BdG}}_{i+1,i}=
\begin{pmatrix}
-t_+&0&0&0\\
0&-t_-&0&0\\
0&0&t_+&0\\
0&0&0&t_-
\end{pmatrix},\quad
\mathcal{H}'^{\mathrm{BdG}}_{i,i+1}=
\begin{pmatrix}
-t_-&0&0&0\\
0&-t_+&0&0\\
0&0&t_-&0\\
0&0&0&t_+
\end{pmatrix},\notag\\
\mathcal{H}'^{\mathrm{BdG}}_{i,i}=
\begin{pmatrix}
-i(t_+-t_-)&\delta h&0&\Delta\\
\delta h&-i(t_+-t_-)&-\Delta&0\\
0&-\Delta&i(t_+-t_-)&-\delta h\\
\Delta&0&-\delta h&i(t_+-t_-)
\end{pmatrix},
\end{align}
where $t_\pm=t\pm\Gamma/4$.

\section{Condition for Infinitesimal instability against local perturbation}
We here discuss conditions for the infinitesimal instability against the local perturbation on the basis of the continuum band introduced in Ref. \cite{YSW-18,Yokomizo-19}.
We also prove that the energy spectrum of our model with the OBC under the infinitesimal perturbation is identical to that with the PBC in the thermodynamic limit.
\subsection{Bulk spectrum under the open boundary condition}
According to Ref. \cite{YSW-18,Yokomizo-19}, the OBC bulk spectrum in the thermodynamic limit is given by the continuum band defined as follows.
Let us consider the generalized Bloch Hamiltonian $H(\kappa)$, where the crystal momentum $k$ is replaced with $\kappa\in\mathbb{C}$ in the Bloch Hamiltonian.
Any energy eigenvalue $E$ of the system satisfties the following characteristic equation:
\begin{align}
\det \left(H(\kappa)-E\right)=0.\label{characteristic}
\end{align}
This is a $2lN$-order algebraic equation of $e^{i\kappa}$ with $l$ and $N$ being the hopping range and internal degrees of freedom per unit cell, respectively.
Let $e^{i\kappa_i} (i=1,...,2lN)$ with $|e^{i\kappa_1}|\le ...\le |e^{i\kappa_{2lN}}|$ be solutions of Eq.($\ref{characteristic}$). Yokomizo and Murakami showed that the OBC bulk band spectrum $E$ is determined so that the relation 
\begin{align}
|e^{i\kappa_{lN}}|=|e^{i\kappa_{lN+1}}|,\label{continuum condition}
\end{align}
holds: Actually, Eq. ($\ref{continuum condition}$) implies
\begin{align}
e^{i\kappa_{lN}}=e^{ik}|e^{i\kappa_{lN+1}}|,\label{momentum}
\end{align}
and thus it gives a one-parameter family of $E$, which is mentioned as the continuum band \cite{Yokomizo-19}.
Note that if $|e^{i\kappa_{lN}}|=|e^{i\kappa_{lN+1}}|=1$, Eq.($\ref{momentum}$) reduces to $e^{i\kappa_{lN}}=e^{ik}$,and the OBC bulk spectrum coincides with the PBC one.
\subsection{Condition for infinitesimal instability against local perturbation}
Note that the above argument should be applied for each sector when the generalized Bloch Hamiltonian is block-diagonalized by a unitary symmetry:
\begin{align}
H(\kappa)=\bigoplus_iH_i(\kappa).
\end{align}
If the continuum bands of two subsystems $i$ and $j$ are degenerated at some region of spectra, then $H_i(\kappa)\oplus H_j(\kappa)$ can be unstable against an infinitesimal perturbation that couples $i$ with $j$.
To see this, let us consider the normal part (without constant dissipation) of our model as the simplest example:
\begin{align}
\sum_{i,\sigma_z=\pm}\left[-t_{\sigma_z}a^\dagger_{i+1,\sigma_z}a_{i,\sigma_z}-t_{(-\sigma_z)}a^\dagger_{i,\sigma_z}a_{i+1,\sigma_z}\right].\label{normalpart}
\end{align}
As explained in the main text, this is a stacked Hatano-Nelson model whose directions of asymmetric hopping terms depend on spin.
The generalized Bloch Hamiltonian for Eq. ($\ref{normalpart}$) is given by
\begin{align}
H(\kappa)=
\begin{pmatrix}
H(\kappa)_\uparrow&0\\
0&H(\kappa)_\downarrow
\end{pmatrix}
:=
\begin{pmatrix}
-t_+e^{i\kappa}-t_-e^{-i\kappa}&0\\
0&-t_-e^{i\kappa}-t_+e^{-i\kappa}
\end{pmatrix}.
\end{align}
By applying the method in Ref. \cite{YSW-18,Yokomizo-19} for each sector, we obtain the conditions for the continuum band ($\ref{continuum condition}$): 
\begin{align}
&|e^{i\kappa^\uparrow_1}|=|e^{i\kappa^\uparrow_2}|=\sqrt{\frac{t_-}{t_+}},\notag\\
&|e^{i\kappa^\downarrow_1}|=|e^{i\kappa^\downarrow_2}|=\sqrt{\frac{t_+}{t_-}}.
\end{align}
These two sectors have degenerated energy $E^{\mathrm{unperturbed}}$, while they have different $|e^{i\kappa}|$, which determines the behavior of the non-Hermitian skin effect.
Next, we consider the local perturbation that can be written in terms of the generalized Bloch Hamiltonian:
\begin{align}
H_\epsilon(\kappa)=
\begin{pmatrix}
H(\kappa)_\uparrow&\epsilon\\
\epsilon&H(\kappa)_\downarrow
\end{pmatrix},
\end{align}
where $\epsilon$ is the perturbation that couples different sectors. In general, the perturbation can be non-Hermitian and take the matrix form.
Since this Hamiltonian is no longer the block-diagonal matrix, and the method in Ref. \cite{YSW-18,Yokomizo-19} should be applied to the whole Hamiltonian to get the continuum band.
For $\epsilon \ll 1$, $E^{\mathrm{unperturbed}}$ is the solution of the characteristic equation that satisfies
\begin{align}
|e^{i\kappa^\uparrow_1}|=|e^{i\kappa^\uparrow_2}|<|e^{i\kappa^\downarrow_1}|=|e^{i\kappa^\downarrow_2}|\label{betaineq}
\end{align}
for $t_+\neq t_-$, but it does not satisfy the condition for the continuum band ($\ref{continuum condition}$), so a completely different energy spectrum realizes.
In the present case, the new energy spectrum turns out to be identical to the one under the PBC by symmertry as shown in the next section.

In summary, the energy spectrum of a non-Hermitian system is unstable if the following condition are satisfied.
\begin{itemize}
\item The generalized Bloch Hamiltonian is block-diagonalized by some unitary symmetry $U$.
\item Two sectors have degenerated energy $E^{\mathrm{unperturbed}}$, while they are characterized by different values of $|e^{i\kappa}|$. 
\item The perturbation that breaks the unitary symmetry $U$ is allowed, and those two sectors are coupled.
\end{itemize}
\subsection{Relationship between symmetry and skin effect}
By combining the method in Ref. \cite{YSW-18,Yokomizo-19} and a symmetry consideration, we can judge whether the non-Hermitian skin effect occurs for certain symmetries.
In our case, the generalized Bloch Hamiltonian obeys 
\begin{align}
\sigma_x\tau_zH_\epsilon(-\kappa)^T\sigma_x\tau_z=H_\epsilon(\kappa).\label{daggersym}
\end{align}
By using this symmetry, we can rewrite Eq. ($\ref{characteristic}$) as
\begin{align}
&\det\left[\sigma_x\tau_z(H_\epsilon(-\kappa)-E)\sigma_x\tau_z\right]=0\notag\\
\Leftrightarrow&\det(H_\epsilon(-\kappa)-E)=0.
\end{align}
Thus, if $e^{i\kappa}$ is a solution of the characteristic equation ($\ref{characteristic}$), then $e^{-i\kappa}$ is also a solution of it.
Owing to this constraint, Eq. ($\ref{continuum condition}$) becomes
\begin{align}
|e^{i\kappa_{lN}}|=|e^{i\kappa_{lN+1}}|=1,
\end{align}
which means that the continuum band is identical to the PBC spectrum with the perturbation $\epsilon$, and the non-Hermitian skin effect does not occurs.
Since we have already taken the thermodynamic limit to discuss the continuum band, the PBC spectrum in $\epsilon\rightarrow0$ limit gives the OBC bulk spectrum in $\lim_{\epsilon\rightarrow0}\lim_{L\rightarrow\infty}$. This proves that the energy spectrum of our model with the OBC under the infinitesimal perturbation is identical to that with the PBC in thermodynamic limit.

Note that the above discussion cannot be applied to the unperturbed Hamiltonian,which can be block-diagonalized. Each sector no longer has the symmetry ($\ref{daggersym}$), while the method in Ref. \cite{YSW-18,Yokomizo-19} is applied not for the whole Hamiltonian but for each sector.  This is the reason why the unperturbed Hamiltonian shows the skin effect.

\section{Creation and annihilation operators of eigenstates in non-Hermitian systems}
We here discuss how the creation and annihilation operators of eigenstates are defined in non-Hermitian systems.
We consider the general quadratic non-Hermitian Hamiltonian
\begin{align}
H=\sum_{i,j}a^\dagger_i\mathcal{H}_{i,j}a_j,\label{quadratic}
\end{align}
where $\mathcal{H}$ is a non-Hermitian Hamiltonian matrix, and $(a,a^\dagger)$ are creation and annihilation operators that satisfy the bosonic or fermionic commutation relations.
Suppose that $\mathcal{H}$ is diagonalizable.
In such a case, physical eigenstates are characterized by the right eigenstates of $\mathcal{H}$.
Let us define the following two matrices by using the right and left eigenstates:
\begin{align}
R:=(|u_1\rangle,|u_2\rangle,\cdots),L:=(|u_1\rangle\!\rangle,|u_2\rangle\!\rangle,\cdots),
\end{align}
where the right and left eigenstates of $\mathcal{H}$ are defined as
\begin{align}
\mathcal{H}|u_n   \rangle=E_n|u_n\rangle,\mathcal{H}^\dagger|u_n   \rangle\!\rangle=E_n^*|u_n\rangle\!\rangle.\label{leftrightdef}
\end{align}
By using these matrices, the biorthonomal condition $\langle m|n \rangle\!\rangle=\langle\!\langle m|n \rangle=\delta_{mn}$ and the completeness condition $\sum_{n}|n\rangle\!\rangle\langle n|=\sum_{n}|n\rangle\langle\!\langle n|$ can be summarized in the following simple form:
\begin{align}
R^\dagger L=L^\dagger R=RL^\dagger=LR^\dagger=1.
\end{align}
Using these relations, $\mathcal{H}$ can be expressed as
\begin{align}
\mathcal{H}=RE L^\dagger=RER^{-1},
\end{align}
where $E=\mathrm{diag}(\cdots,E_n,\cdots)$. Thus, the Hamiltonian ($\ref{quadratic}$) can be rewritten as 
\begin{align}
H=\sum_{n}(\sum_{i}a^\dagger_iR_{i,n})E_n(\sum_{j}R^{-1}_{n,j}a_j)=:\sum_{n}E_n\overline{\alpha}_n\alpha_n.\label{cran}
\end{align}
This is the definition of the creation and annihilation in the new basis.
Although $\overline{\alpha}$ is no longer the Hermitian conjugate of $\alpha$, ($\alpha,\overline{\alpha}$) behave as the creation and annihilation operators that satisfy the bosonic or fermionic commutation relation:
\begin{align}
[\alpha_n,\overline{\alpha}_{n'}]_{\pm}&=R^{-1}_{n,j}R_{i,n'}[a_j,a^\dagger_i]_{\pm}=[R^{-1}R]_{n,n'}=\delta_{nn'},\notag\\
[\alpha_n,\alpha_{n'}]_{\pm}&=[\overline{\alpha}_n,\overline{\alpha}_{n'}]_{\pm}=0,
\end{align}
where $[,]_{\pm}$ denotes the bosonic and fermionic commutation relation. The new vacuum and eigenstates are defined as
\begin{align}
&\alpha_n|0\rangle=0,\notag\\
&|n\rangle=\overline{\alpha}_n|0\rangle.
\end{align}

\section{$\mathbb{Z}_2$ topological invariant of class D point-gapped phase in one dimension}
\begin{figure}[]
\begin{center}
　　　\includegraphics[width=15cm,angle=0,clip]{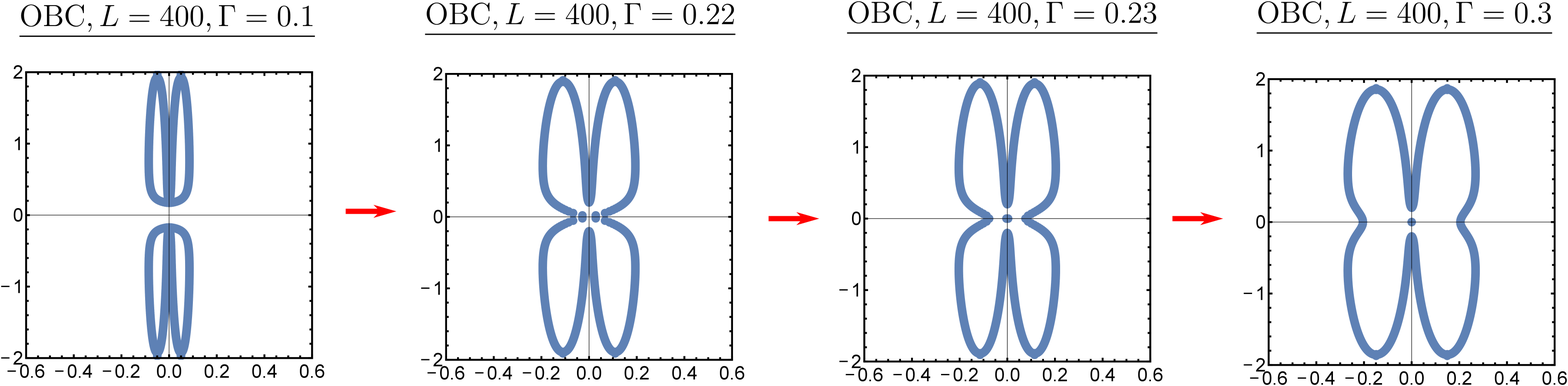}
　　　\caption{Complex energy spectra for various $\Gamma$s. The model parameters are $t_+=t+\Gamma/4=1$, $\Delta=0.2$, $L=400$, and $\delta h=10^{-2}$. The topological phase transition occurs around $\Gamma=0.22$, and Majorana zero modes exist in the nontrivial region.}
　　　\label{supfig}
\end{center}
\end{figure}

We here construct a $\mathbb{Z}_2$ topological invariant in one dimension protected by the particle-hole symmetry
\begin{align}
\tau_x \mathcal{H}^T_k\tau_x=-\mathcal{H}_{-k}.\label{supphsym}
\end{align}
As we noted in the main text, the charge conjugation is defined by using not the complex conjugation but the transpose, and they are inequivalent in non-Hermitian systems.
The point-gap topological classification of the non-Hermitian Hamiltonian is mapped to the topological classification of the corresponding Hermitian Hamiltonian \cite{KSUS-18}:
\begin{align}
\tilde{\mathcal{H}}_k=
\begin{pmatrix}
0&\mathcal{H}_k\\
\mathcal{H}_k^\dagger&0
\end{pmatrix}.\label{mappedhermitian}
\end{align}
The particle-hole symmetry in the mapped Hamiltonian can be written as the conventional antiunitary operation:
\begin{align}
\tau_x\tilde{\mathcal{H}}^*_k\tau_x=-\tilde{\mathcal{H}}_{-k}.
\end{align}
In addition, the Hamiltonian has a chiral symmetry:
\begin{align}
\Sigma_z
\begin{pmatrix}
0&\mathcal{H}_k\\
\mathcal{H}_k^\dagger&0
\end{pmatrix}
\Sigma_z
=-
\begin{pmatrix}
0&\mathcal{H}_k\\
\mathcal{H}_k^\dagger&0
\end{pmatrix}
\ \mathrm{with}\ \Sigma_z:=
\begin{pmatrix}
1&0\\
0&-1
\end{pmatrix}.
\end{align}
Owing to the additional chiral symmetry, the symmetry class is shifted, and the mapped Hamiltonian turns out to be a class DIII Hermitian matrix.
In the Hermitian topological classification, the class DIII topological phases in one dimension are characterized by a $\mathbb{Z}_2$ topological invariant.
Actually, we can construct the $\mathbb{Z}_2$ topological invariant by making use of the off-diagonal basis in Eq. ($\ref{mappedhermitian}$):
\begin{align}
(-1)^\nu=\frac{\mathrm{Pf}(\tau_x\mathcal{H}_{k=\pi})}{\mathrm{Pf}(\tau_x\mathcal{H}_0)}\exp\left(-\frac{1}{2}\int_0^{\pi}dk\mathrm{Tr}[\mathcal{H}_k^{-1}\partial_k\mathcal{H}_k ]\right),\label{invariant}
\end{align}
where $k=0,\pi$ denote the time-reversal invariant points in momentum space. We have used the fact that $\tau_x\mathcal{H}(0)$ and $\tau_x\mathcal{H}(\pi)$ are antisymmetric matrices due to Eq. ($\ref{supphsym}$), and thus the Pfaffian can be naturally defined for them.
Note that the topological invariant ($\ref{invariant}$) is written in terms of the original non-Hermitian Hamiltonian $\mathcal{H}$.
This formula enables us to compute the topological invariant of the model used in the main text:
\begin{align}
\mathcal{H}_{k}=\left[-2t\cos k-\frac{i\Gamma}{2}\right]\hat{\tau}_z-\frac{i\Gamma}{2}\sin k \hat{\sigma}_z-\Delta\hat{\sigma}_y\hat{\tau}_y.\label{supbdg}
\end{align}
Although the analytical expression of the integrand is complicated due to the inverse of the Hamiltonian, we find that the topological invariant for the gapped region ($E\neq0$) is given by
\begin{align}
(-1)^\nu=\mathrm{sgn}\left[|\Delta|-|\Gamma|\right].
\end{align}
Note that the topological phase transition occurs at $|\Delta|=|\Gamma|$, where the point gap of the complex energy band structure of Eq. ($\ref{supbdg}$) is closed.

To check the bulk-boundary correspondence, we perform the numerical diagonalization in real space ($L=400$, OBC) for several $\Gamma$s (Fig.$\ref{supfig}$).
Owing to the slight change of the bulk spectrum that comes form the finite size effect and small but nonzero magnetic filed, the exact correspondence between the finite real-space calculation and the momentum-space one does not hold.
In fact, the phase transition occurs around $\Gamma=0.22$, which differs from $\Delta$ ($=0.2$).
Besides this slight difference, we find that the topological phase transition is clearly accompanied by the near-zero boundary modes.

\section{Majorana condition}
We here discuss the Majorana zero mode in non-Hermitian systems.
Suppose that the Hamiltonian matrix $\mathcal{H}$ has the particle-hole symmetry $C=\tau_x$:
\begin{align}
\tau_x\mathcal{H}^T\tau_x=-\mathcal{H}.
\end{align}
For convenience, we rewrite the theory in the Majorana basis:
\begin{align}
a':=Ua\ \mathrm{with}\ 
U=\frac{1}{\sqrt{2}}
\begin{pmatrix}
1&1\\
-i&i
\end{pmatrix},\quad a'^\dagger=a'.
\end{align}
In this basis, the particle-hole symmetry$C$ is equal to unity:
\begin{align}
&a^\dagger \mathcal{H} a=a'^\dagger [U\mathcal{H}U^{-1}] a'=:a'^\dagger \mathcal{H'} a',\notag\\
-\mathcal{H}&=\tau_x\mathcal{H}^T\tau_x=\tau_xU^T[(U^T)^{-1}\mathcal{H}^TU^T](U^T)^{-1}\tau_x\notag\\
\Leftrightarrow -\mathcal{H}'&=[U\tau_xU^T]\mathcal{H}'[(U^T)^{-1}\tau_xU^{-1}]\notag\\
&=\mathcal{H}'^T,
\end{align} 
where we have used the explicit form of $U$ and $\tau_x$ in the last line.
In the following, we use the Majorana basis and omit $'$.

In the Majorana basis, the following equation holds:
\begin{align}
\mathcal{H}^\dagger|u_n   \rangle\!\rangle&=E_n^*|u_n\rangle\!\rangle\notag\\
\Leftrightarrow\mathcal{H}^T|u_n   \rangle\!\rangle^*&=E_n|u_n\rangle\!\rangle^*\notag\\
\Leftrightarrow\mathcal{H}|u_n   \rangle\!\rangle^*&=-E_n|u_n\rangle\!\rangle^*.
\end{align}
We have used the definition of the left eigenfunction ($\ref{leftrightdef}$) in the first line and $\mathcal{H}^T=-\mathcal{H}$ in the last line.
Thus, the particle-hole symmetry ensures the existence of the presence of eigenfunction with $-E_n$ for each eigenfunction with $E_n$ except for the zero mode.
If the number of zero mode is one, then 
\begin{align}
|u_0\rangle=|u_0\rangle\!\rangle^*.\label{leqr}
\end{align}
By using Eq. ($\ref{cran}$), we can write the creation and annihilation operators for the zero mode as
\begin{align}
\overline{\gamma}=|u_0\rangle_ia_i^\dagger,\quad
\gamma=|u_0\rangle\!\rangle^*_ia_i.
\end{align}
Using the property of the Majorana basis $a_i^\dagger=a_i$ and Eq. ($\ref{leqr}$), we obtain the Majorana condition:
\begin{align}
\overline{\gamma}=\gamma.
\end{align}

%